\documentclass[aps,prb,twocolumn,floatfix,amsmath,amssymb]{revtex4}
\usepackage{graphicx}
\usepackage{dcolumn}
\usepackage{bm}
\usepackage{epsfig,psfrag,amsmath,amssymb,float}
\input{epsf}
\newcommand{\vv}[1]{{\bf #1}}

\begin{document}

\title{Renormalization of the electron-spin-fluctuation interaction 
in the $t-t'-U$ Hubbard model}
\author{Z.~B.~Huang$^{1,2}$, W.~Hanke$^1$, E.~Arrigoni$^{3}$, and
A.~V.~Chubukov$^4$}
\affiliation{$^1$Institut f\"ur Theoretische Physik, 
Universit\"at W\"urzburg,
am Hubland, 97074 W\"urzburg, Germany}
\affiliation{$^2$Department of Physics, Hubei University, Wuhan 430062, PRC}
\affiliation{$^3$Institut f\"ur Theoretische Physik, 
Technische Universit\"at Graz, Petersgasse 16, A-8010 Graz, Austria}
\affiliation{$^4$Department of Physics, University of Wisconsin, Madison, 
Wisconsin 53706, USA}
\date{\today}

\begin{abstract}
We study the renormalization of the 
electron-spin-fluctuation (el-sp) vertex in a two-dimensional 
Hubbard model with nearest-neighbor ($t$) and next-nearest-neighbor
($t'$) hopping by a Quantum-Monte-Carlo calculation.
We distinguish between el-sp vertices involving interacting particles
and quasiparticles, i.e. separate the renormalizations of 
the vertex from that of the quasiparticle residue $1/Z$. 
We show that for $t'=0$, 
the renormalized el-sp vertex, not dressed by $1/Z$, decreases 
with decreasing temperature at all momentum transfers. 
As a consequence, the effective pairing interaction 
mediated by antiferromagnetic spin fluctuations is reduced due to 
vertex corrections. The inclusion of a 
finite $t'/t<0$, increases the Landau damping rate 
of spin fluctuations, especially in the overdoped region.
The increased damping rate 
leads to smaller vertex corrections, in agreement 
with earlier diagrammatic calculations.  
Still, corrections reduce the spin-fermion vertex even at
finite $t'$.
\end{abstract}

\maketitle
\section{Introduction}
The nature of the spin-fluctuation mediated interaction between charge 
carriers 
continues to attract high interest in the high-T$_c$ 
superconductivity research~\cite{scalapino,IFT98,assaad,norman,carbotte,chub}.
Theoretical calculations for the doped Hubbard
and $t-J$ models suggest that the 
strongest interaction between fermions is due to antiferromagnetic (AF) 
spin fluctuations. The low-energy description of this interaction 
has been advanced in the framework of the spin-fermion model~\cite{chub}.
A magnetically mediated interaction has been intensively studied 
since the early days of high-$T_c$ era because AF spin fluctuations 
induce $d-$wave pairing between fermions~\cite{scalapino}.
Various numerical and analytical calculations based on 
Hubbard and $t-J$ models, as well as semi-phenomenological spin-fermion 
models, 
have been performed in the normal and superconducting states of the cuprates. 
The results of these calculations are in quite good agreement with a
large number of  experimental data. 
In particular, these calculations show that the magnetic 
response in a $d-$wave superconducting state contains a resonance peak, 
in addition to a gapped continuum. The location of the peak, and 
its ``negative'' dispersion are in agreement with experimental data. 
It has been argued~\cite{finger}  that the scattering from the 
resonance is strong enough to explain the peak-dip-hump feature in 
the fermionic spectral function, the $S-$shape dispersion for 
antinodal fermions, and 
similar features in SIS tunneling, optical conductivity, 
and Raman response.  

The strength of the feedback effects from the resonance peak on fermions, and 
the magnitude of the spin-mediated $T_c$ depend on the size of the spin-fermion 
coupling. The estimates for the coupling strength still 
vary substantially. 
Abanov et al argued~\cite{abanov} that the coupling $g$ 
is comparable to the Hubbard $U$ and is quite strong, $g \sim 0.7 eV$.  
On the other hand, Kee {\it et al.}~\cite{kee} argued that the effective 
spin-fermion coupling  is much weaker $g \sim 0.014 eV$.
In the latter work~\cite{kee},  the coupling was extracted 
from the specific heat data.  To a large extent, the 
difference between the two results  is related to 
different choices for the fermionic density of states $N_0$: 
a large value of $g$ is obtained by assuming a large Luttinger Fermi surface, 
with $N_0 \sim 1 eV^{-1}$ (Ref.~\onlinecite{abanov}), 
while a smaller $g$ is obtained 
by assuming that the density of states $N_0 \sim J^{-1} \sim 10 eV^{-1}$ is 
the same as in a weakly doped quantum AF. 

This difference brings in the issue of how much the full spin-fermion vertex 
$g_{\vv k\vv q}$  differs from the bare $g^{0}_{\vv k\vv q}$ due to vertex corrections. 
Here, $\vv k$ and $\vv k + \vv q$ are incoming and outgoing 
fermionic momenta, and $\vv q$ is the 
bosonic momentum.
The ratio $\Gamma(\vv k, \vv q) 
= g_{\vv k\vv q}/g^{0}_{\vv k\vv q}$ determines the vertex
renormalization originating from
electronic correlations. 
Some of us have recently shown numerically that this quantity is substantially
renormalized by strong electronic correlations
both for the electron-phonon
(el-ph) vertex~\cite{zbhuang_ph} as well as for the el-sp vertex~\cite{zbhuang_sp}.

Within the spin-fermion model,
the role of the vertex renormalization is well understood 
in the limits of very small and relatively large doping. 
At very small doping $x$, when long-range AF
order is still present, nearly all carriers are localized, and doped fermions 
form small pockets around $(\pi/2,\pi/2)$ and symmetry related points. 
In this situation, the full spin-fermion vertex $g_{\vv k\vv q}$ vanishes 
when $\vv q$ coincides with the antiferromagnetic momentum 
$\vv Q \equiv (\pi,\pi)$. The vanishing of $g_{\vv k\vv Q}$ is the 
consequence of the Adler principle that the interaction should preserve
massless  Goldstone bosons  in the theory and, therefore, should vanish 
at the ordering momentum of Goldstone bosons. For $\vv q \neq \vv Q$, 
the vertex is non-zero, but small in $|\vv q - \vv Q|/|\vv Q|$. 
Mathematically, the vanishing of $g_{\vv k\vv Q}$ is the result of 
dressing up of the bare interaction $g^0_{\vv k\vv Q}$ by coherence 
factors, associated with the antiferromagnetic order.  From this perspective, 
the strong reduction of $g_{\vv k\vv q}$ when $\vv q$ is close to 
$\vv Q$ is the result of a strong vertex renormalization. Schrieffer
argued~\cite{schrieffer}
that, if the pocket-like Fermi surface survives 
even when long-range AF order is lost,
the strong vertex renormalization extends into the paramagnetic phase.
According to Ref.~\onlinecite{schrieffer} 
\begin{equation}
\Gamma^2(\vv k,\vv q)\propto [(\vv q - \vv Q)^{2}+
\frac{1}{\xi^{2}}]
\sim \frac{1}{[\chi(\vv q,\omega=0)]},
\label{gamma_sc}
\end{equation}
where $\xi$ is the AF correlation length. Once Eq. (\ref{gamma_sc}) is valid,
the effective spin-mediated interaction 
\begin{equation}
V(\vv k,\vv q)\propto |\Gamma(\vv k,\vv q)|^{2}
\chi(\vv q),
\label{pair}
\end{equation}
is considerably
weaker than without 
vertex renormalization, and, most importantly, it is no longer peaked at 
$\vv q\sim \vv Q$.   
This is an important fact, since the peak structure at 
$\vv q\sim \vv Q$
was responsible for a strong attractive $d-$wave 
component in the pairing potential. The magnetically-mediated $d-$wave
pairing is still possible even in this situation,
but the corresponding $T_c$ is 
much smaller than one would get without vertex corrections~\cite{sushkov}. 

In the opposite limit 
of large doping, fermions are itinerant, and have a large Fermi 
surface which crosses the magnetic Brillouin zone boundary at hot spots 
(this requires a sufficiently strong next-nearest-neighbor hopping).
A spin-fluctuation with momentum near $\vv Q$ then can decay into a particle 
and a hole. 
Analytical calculations within the spin-fermion model 
show that in this situation,  the vertex is only weakly renormalized, 
and actually increases with respect to its bare value, i.~e.
$g_{\vv k\vv Q} > g^0_{\vv k\vv Q}$~\cite{chub,aim}. 
More precisely,
the relation between 
$g_{\vv k\vv Q}$ and $g^{0}_{\vv k\vv Q}$ depends on the angle $\phi\leq\pi$ 
between Fermi velocities at hot spots separated by $\vv Q$~\cite{chub}:
\begin{equation}
\Gamma (k, Q) = \frac{g_{\vv k\vv Q}}{g^0_{\vv k\vv Q}} = 
 \left(1 +  \frac{\phi}{4\pi}~ \log{\xi/\xi_0}\right) ,
\label{a_1}
\end{equation}
where 
$\xi_0$ is of order 1.
Equation~(\ref{a_1}) shows that
the renormalization is largest when $\phi \rightarrow \pi$. This
limit corresponds to an almost nested Fermi surface at hot spots.
In optimally doped cuprates, the velocities at the hot spots separated by 
$\vv Q$ are almost perpendicular to each other, i.~e., $\phi \approx \pi/2$. 
For $\xi \sim 2$, the vertex correction is then about 10\% (even smaller, 
$\sim 4$\%, when one adds regular terms~\cite{chub2,chub3}).

The difference between the itinerant and localized limits raises 
the question how vertex corrections evolve as the system approaches 
half-filling. One attempt to address this issue was 
undertaken in Ref.~\onlinecite{chub3}. 
The authors of Ref.~\onlinecite{chub3} 
 considered a toy spin-fermion model in which one can study
the evolution from the Luttinger Fermi surface towards hole pockets. 
This evolution involves a topological transition below which a pocket 
is splitted from a large Fermi surface, and the Luttinger theorem breaks down. 
Within this model, vertex corrections
evolve together with the Fermi surface: they are small and positive 
(i.~e. $g_{\vv k\vv Q}  > g^0_{\vv k\vv Q}$)
in the limit when the Fermi surface is large, but change their sign near the
topological transition, become negative and rapidly increase in magnitude 
as the Fermi surface evolves towards hole pockets. For a pocket-like Fermi 
surface, vertex corrections almost cancel the bare vertex, and the full 
$g_{\vv k\vv q}$ coincides with Schrieffer's result, Eq. (\ref{gamma_sc}). 

In this paper, we address the issue of the strength of  vertex 
corrections in the two-dimensional $t-t'-U$ Hubbard model by means of 
a quite accurate numerical method -- the determinantal
Monte-Carlo~\cite{BSS81} algorithm. The purpose of our calculation is
to explore the extent of validity in doping and temperature of 
Schrieffer's argument~\cite{schrieffer} as well as of the
spin-fermion model calculation~\cite{chub}: Whereas the analytical results 
are only valid when the correlation length $\xi$
is large, Monte-Carlo calculations are not restricted by this condition.
We also want to get an insight as up to what extent 
a description in terms of low-energy spin degrees of freedom, i.~e. in
terms of a spin-fermion
model is appropriate for the  Hubbard model.

Vertex corrections have been analyzed by some of us 
in Ref.~\onlinecite{zbhuang_sp}. The present paper extends this
earlier letter in three different directions. First,
in order to separate vertex
and quasiparticle renormalization effects, we analyze 
the vertex renormalization and the pairing interaction with and
without the inclusion of the wave-function renormalization $Z$.
The vertex without the renormalization
of $Z$ is labeled by $\Gamma(p,q)$ before and in 
Eq.~(\ref{five}) below.  This  vertex is 
for the interaction between spin fluctuations and strongly interacting 
particles. This vertex has to be used for comparison
with analytical results which don't include the rescaling by the 
fermionic $Z$. 
The fully renormalized vertex is the one  rescaled by $Z$  
(we label it by $\gamma(p,q)$, see Eq.~(\ref{nine}) below)~\cite{note1}. 
It describes the interaction between spin fluctuations and quasiparticles.
We discuss the difference between the two vertices in 
Sec.~\ref{Results}, and also discuss there the behavior of the pairing
interaction between interacting particles and quasiparticles. 
Second,  we calculate the momentum dependence of
the vertex. Third,  we discuss in detail 
the effect of bosonic damping on the strength of the vertex renormalization.
We show  that the introduction of a
next-nearest-neighbor hopping $t'$, which makes the bosonic damping possible, 
substantially reduces the  renormalization of $\Gamma$ (but not of $\gamma$).

Our numerical Monte Carlo simulations are performed 
on a $8 \times 8$ lattice at different doping densities and 
different temperatures. In our finite lattice, the 
$(\pi,0)$ point is the one closest to the
 hot spots, so that   charge carriers 
near the $(\pi,0)$  region are strongly affected by antiferromagnetic spin 
fluctuations. Therefore, we will mainly concentrate on the particular scattering
process in which the incoming electron and spin fluctuation carry momenta 
${\vv p}=(-\pi,\,0)$ and ${\vv q}=(\pi,\,\pi)$, respectively.
Within our $\vv p$-points mesh, the points $\vv p$ and $\vv p+ \vv q$ 
lie sufficiently close to the Fermi surface.

Our paper is organized as follows: In Section~\ref{Model}, 
we define the Hamiltonian and describe the numerical approach, which
is based on the Quantum-Monte-Carlo 
evaluation of the linear response to an external spin perturbation.
In Section~\ref{Results}, we present our numerical results and discuss
their qualitative relation with the spin-fermion-model calculation and 
Eq.~\ref{gamma_sc}.
Finally, in Section~\ref{Conclusions}, we discuss in detail 
our main conclusions.

\section{\label{Model} Model and numerical approach}
Our starting point is the one-band Hubbard model,
\begin{eqnarray}
\label{ham}
 H&=&-t \sum_{\langle ij \rangle,\sigma}
     (c_{i\sigma}^\dagger c_{j\sigma}^{\,}
     +c_{j\sigma}^\dagger c_{i\sigma}^{\,})
    -t' \sum_{\langle\langle ij \rangle,\sigma}
     (c_{i\sigma}^\dagger c_{j\sigma}^{\,}
     +c_{j\sigma}^\dagger c_{i\sigma}^{\,})\nonumber\\
     &&+ U \sum_i n_{i\uparrow}n_{i\downarrow},
\end{eqnarray}
Here, the operators $c_{i\sigma}^\dagger$ and $c_{i\sigma}^{\,}$  as usual create 
and destroy an electron with spin $\sigma$ at site $i$, respectively, and 
$\langle ij\rangle$ and $\langle\langle ij\rangle\rangle$
denote a sum over nearest 
and next-nearest neighbor lattice sites $i$ and $j$, respectively.
$n_{i\sigma}= c_{i\sigma}^\dagger c_{i\sigma}^{\,}$ is the number operator.
Finally, $U$ is the onsite Coulomb interaction and the 
nearest-neighbor hopping $t$ is chosen as the unit of energy.
In order to consider the effects of damping, we include a
next-nearest-neighbor hopping term $t'$.

\begin{figure}
\centering
\epsfig{file=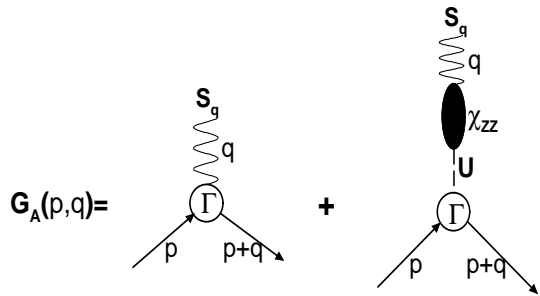,height=7cm,width=4cm,angle=270}
\caption{Diagrammatic representation of $G_A(p,q)$
within linear response to $S_{\vv q}$. The thick solid lines
represent dressed single-particle Green's functions of
the Hubbard model. The wavy line denotes the external
perturbation in Eq.~(\ref{el-sp}). The dashed line represent
the Hubbard interaction $U$ and the black ellipse stands for
the longitudinal spin susceptibility $\chi_{zz}(q)$.}
\label{svertex}
\end{figure}

In our simulation, we use a linear-response 
approach (see also Ref.~\onlinecite{zbhuang_ph})
in order to extract the el-sp vertex.
In this method, one formally adds to Eq.~\eqref{ham} the interaction
with a momentum- and (imaginary) time-dependent spin-fluctuation field
in the $z$-direction
$S_{ \vv q} e^{-i q_0 \tau}$~\cite{pq} in the form~\cite{pq}
\begin{equation}
\label{el-sp}
H_{\rm el-sp}= \sum_{\vv k \vv q\sigma} g_{\vv k\vv q}^{0}
\sigma c_{\vv k+\vv q\sigma}^{\dagger}c_{\vv k\sigma} \
S_{ \vv q} \ e^{-i q_0 \tau}\;,
\end{equation}
where $g_{\vv k\vv q}^{0}$ is the bare el-sp coupling (equal to
the Hubbard $U$ in the one-band Hubbard model).
One then considers the
``anomalous'' single-particle propagator in the presence of this
perturbation defined as~\cite{pq}
\begin{equation}
\label{gq}
G_{A}(p,q)\equiv 
- \int_{0}^{\beta}d\tau\ e^{i(p_{0}+q_{0}) \tau}
 \langle T_{\tau}c_{\vv p+\vv q\sigma}(\tau)c_{\vv p\sigma}^{\dagger}
(0)\rangle_{H+H_{\rm el-sp}},
\end{equation}
Here $\langle\rangle_{H+H_{\rm el-sp}}$ is the Green's function
evaluated with the Hamiltonian $H+H_{\rm el-sp}$.
Diagrammatically, $G_A(p, q)$ has the structure shown in
Fig.~\ref{svertex}, so that the el-sp vertex $\Gamma(p,q)$
can be expressed quite generally in terms of $G_A$ and
of the single-particle Green's function $G(p)$ in
the form
\begin{equation}
\Gamma (p, q) = \lim_{S_{\vv q}\to 0}\frac{1}{g_{kq}^{0}} 
\frac{1}{S_{\vv q}}
\frac{1}{1+U\, \chi_{zz}(q)}\frac{ G_A (p, q)}{G(p+q)\, G (p)} \;,
\label{five}
\end{equation}
with $\chi_{zz}(q)$ the longitudinal spin susceptibility.
Because only the limit $S_{\vv q}\to 0$ is relevant in Eq.~\ref{five},
it is sufficient to calculate the leading 
linear response of $G_A$ to $H_{\rm el-sp}$, which is given by
\begin{eqnarray}
G_A(p,q) = S_{\vv q} \
 \int_{0}^{\beta} d\tau e^{i(p_{0}+q_{0})\tau}
\int_{0}^{\beta} d\tau^{'} e^{-i q_{0} \tau'}
\sum_{\vv k\sigma^\prime} g_{\vv k\vv q}^{0}
\times \nonumber\\
\langle T_{\tau}\sigma^\prime c_{\vv k+\vv q\sigma^\prime}^{\dagger}
(\tau'+0^{+})c_{\vv k\sigma^\prime}(\tau')
 c_{\vv p+\vv q\sigma}(\tau)c_{\vv p\sigma}^{\dagger}(0)\rangle_{H},
\label{four}
\end{eqnarray}
where $0^+$ is a positive infinitesimal. Notice that $S_{\vv q}$
cancels in Eq.~\ref{five}.
The two-particle Green's function in Eq.~\eqref{four} is evaluated
with respect to the pure Hubbard Hamiltonian  (Eq.~\eqref{ham}).

Within a quasiparticle approach, one can shift the wave-function
renormalization factor
 $Z(p) >1$ from the Green's function into the definition of 
the effective el-sp vertex.~\cite{zbhuang_ph} 
Thus, for states close to the Fermi energy with a single-particle
Green's function of the form $G(p)=Z(p)^{-1} \ (i\ p_0-E_p)^{-1}$,
where $E_p$ is the quasiparticle energy, the effective quasiparticle-spin
coupling is given by:
\begin{equation}
\gamma(p, q) = \frac{\Gamma(p, q)}{\sqrt{Z(p)\, Z(p+q)}}\;.
\label{nine}
\end{equation}
Numerically, $Z(p)$ is evaluated  as 
$Z(p)={\rm Im}[1/G(p)]/p_0$~\cite{zbhuang_ph,pq}.
Therefore, $\gamma$ is the vertex between {\em
 quasiparticles} and spin fluctuations. 

The total pairing interaction
$V_p$ for the 
exchange of a single spin fluctuation can be expressed in terms of the
vertex $\Gamma$ as
\begin{equation}
V_{p}(p,q)=|\Gamma(p,q)|^{2}\cdot U^{2}\cdot\chi_{zz}(q) \;,
\label{totalp}
\end{equation}
where 
$\chi_{zz}(q)$ is
the spin susceptibility 
\begin{eqnarray}
\chi_{zz} (q) & = &\frac{1}{2}\int^\beta_0 d\tau\ e^{-i\ q_0 \tau}
\ \left\langle T_\tau m_{\vv q}^{z} (\tau) m_{-\vv q}^{z} (0)
\right\rangle,
\nonumber\\
\noalign{\hbox{and}}
m_{\vv q}^{z} & = & {1\over\sqrt{N}}\sum_{\vv k\sigma}
\sigma c_{\vv {k+q}\sigma}^{\dagger}c_{\vv k\sigma}\;.
\label{ten}
\end{eqnarray}
Eq.~\ref{totalp} describes the effective pairing interaction 
between {\it interacting particles}. Including the wave-function renormalization
as in Eq.~\ref{nine}, we introduce the effective pairing interaction 
{\it between 
quasiparticles} 
(this is in complete analogy to the treatment of the strong-coupling
superconductors in Ref.~\cite{scal.69})
\begin{equation}
v_p (p,q) = \frac{ V_p(p,q) }{Z(p) \ Z(p+q)}  = |\gamma(p,q)|^2 U^2 \chi_{zz} (q)\;.
\label{effectivep}
\end{equation}
Clearly, both
$V_{p}(p,q)$ and $v_p (p,q)$ contain the contributions from 
both the real and imaginary parts of the vertex.
Within a Quantum-Monte-Carlo approach,
the effects and strength of the particle-particle vertex in a given
channel can be also indirectly 
measured by comparing the full pairing susceptibility
with the corresponding ``bubble'' approximation~\cite{wh.sc.89}.

\section{\label{Results} Results and Discussions}
\begin{figure}
\centering
\epsfig{file=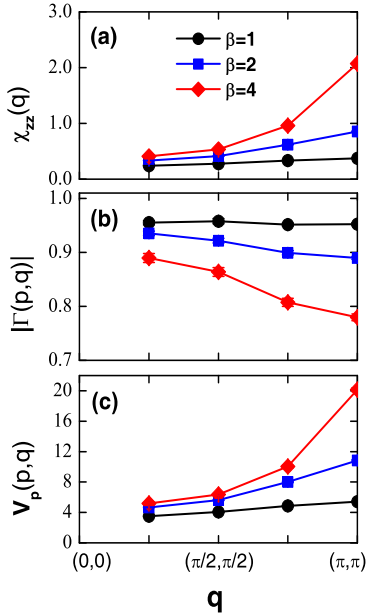,height=9cm,width=5.5cm,angle=0}
\caption{(color online) 
(a) Spin susceptibility $\chi_{zz}(q)$, (b) the renormalized
el-sp vertex $|\Gamma(p,q)|$, and (c) the total pairing interaction
$V_{p}(p,q)$ as a function of spin-fluctuation momentum transfer $\vv q$.
Here $U=4$, $t'=0$, ${\vv p}=(-\pi,0)$, and the doping density 
$\delta=0.12$. 
The value of the inverse temperature $\beta$ is indicated
by the shape of the symbol.}
\label{Vp_Gamma_chi}
\end{figure}
In order to explore 
the momentum structure of the el-sp interaction, 
we first plot $\chi_{zz}(q)$, $|\Gamma(p,q)|$,
and $V_{p}(p,q)$ versus $\vv q$ at different temperatures in
Fig.~\ref{Vp_Gamma_chi}. 
Here, the spin-fluctuation
momentum transfer $\vv q$ is along the $(1,1)$ direction and $U$ is at
an intermediate coupling value, i.e. $U=4$. One can readily see that both
$\chi_{zz}(q)$ and $V_{p}(p,q)$ are peaked at momentum transfers 
around the antiferromagnetic vector $\vv Q = (\pi,\pi)$,
and the
strength of the peaks increases when the temperature is lowered. This demonstrates 
that the $\vv q$- and $T$-dependences of the spin susceptibility $\chi$ dominate
the temperature behavior of the pairing interaction, despite the
reduction of the vertex $\Gamma$ as the temperature is decreased.
From Fig.~\ref{Vp_Gamma_chi}, we observe that 
the decrease with temperature 
of $\Gamma$ at large momentum transfers is stronger than at small momentum
transfers, indicating
that the vertex correction at $\vv q\sim \vv Q$ is larger than
at small $\vv q$.
This $\vv q$-dependence of the el-sp vertex is qualitatively in
agreement with the prediction of Eq.(\ref{gamma_sc}). Our finding is also in good 
agreement with the work of Bulut {\it et al.}~\cite{bulut} (carried
out at $U=4$), which shows 
a value of $g=0.8$ (corresponding $|\Gamma|=0.8$ in our notation) can produce
an effective coupling $gU$ which is consistent with the results of Monte Carlo
calculation of the irreducible particle-hole vertex, and that the effective 
particle-particle interaction originating from the Hubbard $U$ increases with 
lowering temperature and can reach large values.
\begin{figure}
\centering
\epsfig{file=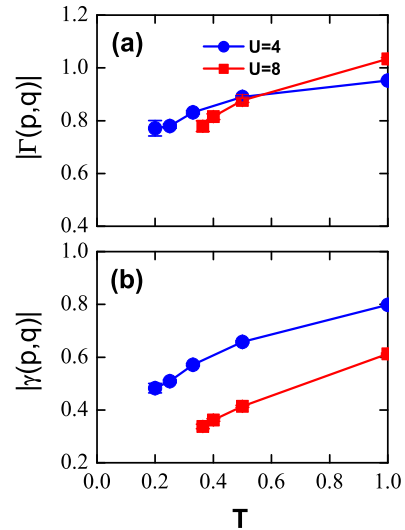,height=8cm,width=6cm,angle=0}
\caption{(color online)
The absolute values of the (complex) renormalized 
el-sp vertices $|\Gamma(p,q)|$ and $|\gamma(p,q)|$ vs $T$ 
at $U=4$ and $U=8$ at the doping density
$\delta=0.12$. We set ${\vv p}=(-\pi,0)$, ${\vv q}=(\pi,\pi)$, and $t'=0$.}
\label{Gam_gam_T}
\end{figure}

\begin{figure}
\centering
\epsfig{file=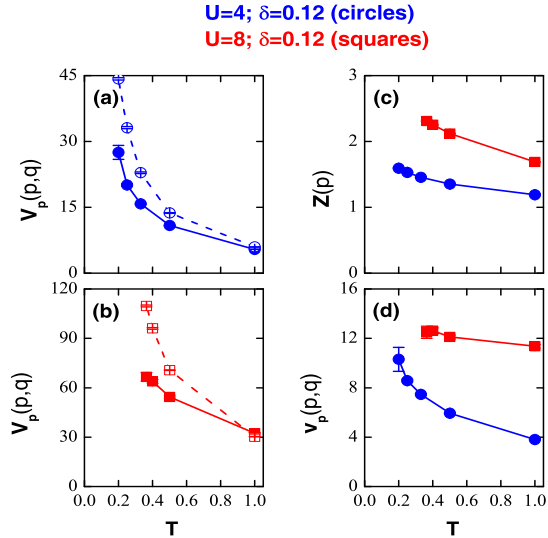,height=8cm,width=8cm,angle=0}
\caption{(color online) 
The pairing interaction $V_{p}(p,q)$ vs $T$ 
for (a) $U=4$ and (b) $U=8$ at the doping density $\delta=0.12$ (full
symbols, the solid line is a guide to the eye). 
The open symbols (dashed line) show the RPA results.
In panel (c), we show the $T$-dependence of the wave-function 
renormalization $Z(p)$ and in panel (d) 
the effective pairing interactions $v_p (p,q)$ between quasiparticles. 
In all panels,
 ${\vv p}=(-\pi,0)$, ${\vv q}=(\pi,\pi)$, and $t'=0$. }
\label{VpV_Z_T}
\end{figure}
Monte Carlo results for $\Gamma(p, q)$, $\gamma(p,q)$, 
$Z(p)$, $V_{p}(p,q)$ and $v_p (p,q)$ are displayed 
in Figs.~\ref{Gam_gam_T} and \ref{VpV_Z_T} for intermediate as well
as strong correlation.  
We compute the vertices at zero bosonic Matsubara frequency and 
the smallest fermionic Matsubara frequency $\omega_1 = \pi T$.  
Because $\omega_1$ is finite, and also because at a nonzero doping,
the excitation spectrum is particle-hole asymmetric, the vertices 
$\Gamma$ and $\gamma$ have both real and imaginary parts 
(the imaginary part obviously vanishes at $T=0$).  
We notice that in the underdoped regime ($\delta=0.12$)
for both intermediate  ($U=4$) and strong correlation 
($U=8$), both $|\Gamma|$ and $|\gamma|$ are strongly renormalized 
below a characteristic temperature ($T\lesssim J=0.5$ for $U=8$).
Although our simulation cannot reach low temperatures,
a clear trend is observable in both $|\Gamma|$ and
$|\gamma|$, which tend to go to small values at low temperatures, 
at least for $U=8$. A more careful look shows that the reduction of the
 vertices $\Gamma$ and $\gamma$ is chiefly due to the reduction of 
$Re \Gamma$ and $Re \gamma$. The imaginary parts of the vertices 
are small at weak and intermediate correlations (at our lowest accessible $T$),
but for $U=8$ they can 
become comparable to the real parts. 

Further, our  numerical results 
presented in Fig.~\ref{VpV_Z_T}
clearly 
show that, in both intermediate- and strong-correlation regimes,
the renormalized pairing interaction $V_{p}$ (which does not include 
the dressing by the quasiparticle $Z$) is smaller than the RPA
result obtained with the full susceptibility $\chi_{zz}$. 
The difference becomes strongest 
 at the lowest
accessible temperatures. 
The qualitative temperature behavior 
of the effective pairing interaction between quasiparticles $v_p$, 
is quite similar to $V_p$ in the 
intermediate-correlation ($U=4$) regime. However, it is quite 
different in the strong-correlation ($U=8$) regime, where 
it displays a mild increase
or a saturation at low $T$'s. 
Based on the results
shown in Fig.~\ref{VpV_Z_T}, we conclude 
that for $t^{'} =0$, both vertex corrections and the renormalization
of $Z$ are 
important for the spin-mediated
 pairing interaction.  
Vertex corrections  suppress $V_p (p,q)$ 
compared to the case of free fermions. The rescaling by $Z$ further
 suppresses the effective pairing interaction between 
quasiparticles, $v_p (p,q)$. Note, though, that the pairing interaction 
$V_p (p,q)$ at a bosonic $q= (\pi,\pi)$ and fermionic $p$ near the 
antinodal point still increases with decreasing temperature.

Fig.~\ref{VpV_Z_T} shows that vertex corrections are substantial
even in the case when
the AF correlation length $\xi$ is quite
small~\cite{groeber}. i.~e. 
of order of the Cu-Cu distance. This is in contrast to 
the situation discussed by
Schrieffer in which holes move in an
AF background, which is unaffected
by the charge carriers, i.~e. $\xi$ is large.
How can we understand this result?
In the AF precursor (large-$\xi$) case~\cite{schrieffer,chub3},
the vanishing of the el-sp interaction in the long-range ordered AF state 
is the result of dressing up of the bare interaction by the AF 
coherence factor, which is small at the top of the 
valence band. 
The coherence factor is due to the interference 
effect of the quasiparticle, which forms a ``spin-bag'', i.~e. 
a hole dressed by a short-range AF background. 
Although in our calculation the AF precursor is no longer present, it is 
well established from our earlier QMC work on the evolution of the 
single-particle spectral 
function $A({\vv k},\omega)$~\cite{groeber} from an insulator to a 
metal that below $T\sim J$, the electronic excitations display an 
essentially doping-independent feature. More precisely, 
a ``band'' of width $J$ forms, in which ``spin-bag''-like quasiparticles 
propagate coherently. The continuous evolution can be traced back to
one and the same many-body origin: the doping-dependent AF spin-spin
correlation. Therefore, we argue that the interference effect
of the spin-bag quasiparticle plays a similar role in reducing the el-sp vertex 
in the strongly-correlated underdoped regime.

\begin{figure}
\centering
\epsfig{file=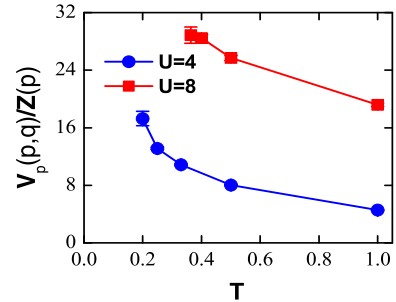,height=5cm,width=6cm,angle=0}
\caption{(color online) $V_{p}(p,q)/Z(p)$ vs $T$ 
for $U=4$ and $U=8$ at the doping density $\delta=0.12$ (full
symbols, the solid line is a guide to the eye). 
${\vv p}$ and ${\vv q}$ are the same as in Fig.~\ref{VpV_Z_T}.}
\label{Pair_VZ_T}
\end{figure}
 
Above we considered the pairing vertices for 
interacting particles and for quasiparticles. 
The actual strength of the pairing interaction is 
proportional to the product of the effective 
interaction $v_p$ and the density of states at the Fermi level $N(E_F)$. 
Since we restricted with only the pole component of $G$, the conservation 
of the particle number implies that $N(E_F)$ is increased by a factor 
$Z$ with respect to the
noninteracting density of states $N_0(E_F)$, i.e., $N (E_F)  v_p =
N_0(E_F) \ V_p /Z$. This is similar to the 
case of the electron-phonon mediated pairing, where the interaction is 
also rescaled by one power of $Z$ (this last rescaling is a well-known 
McMillan result $\lambda \rightarrow \lambda/(1+ \lambda)$. 
Our consideration, in which we defined $Z$ as the overall wave-function
renormalization is formally different from electron-phonon problem, 
where the fermionic self-energy depends only on frequency.  However,
the final result is the same as in the case where self-energy only 
depends on frequency, the density of states is not renormalized by $Z$,
but one power of $Z$, compared to $v_p \sim V_p/Z^2$ is eliminated 
by the mass renormalization $m^*/m = Z$. 

In order to have a measure of the pairing strength including the
renormalization of the density of states, we thus plot the quantity
$V_p(p,q)/Z(p)$ in Fig.~\ref{Pair_VZ_T} as a function of
temperature for $\delta=0.12$. 
We observe that $V_p(p,q)/Z(p)$ increases with decreasing
temperature in both intermediate- and strong-correlation regimes.
On the other hand, the suppression
of $V_{p}$ due to vertex corrections and the increase of $Z$ with
decreasing temperature
considerably reduce the pairing  (i.~e., without
vertex corrections),
in particular in the low-temperature regime.

\begin{figure}
\centering
\epsfig{file=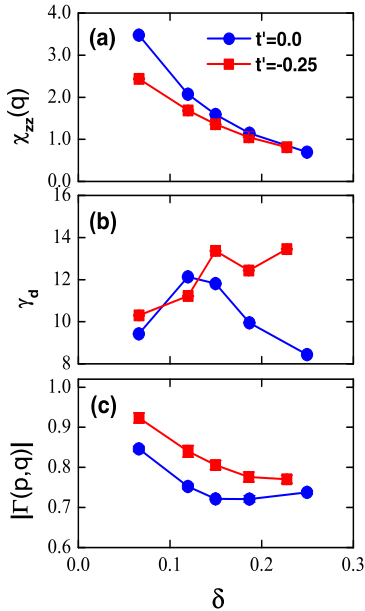,height=9cm,width=5.5cm,angle=0}
\caption{(color online)
(a) Spin susceptibility $\chi_{zz}(q)$, (b)  ``damping rate'' $\gamma_d$, 
and (c)  el-sp vertex $|\Gamma(p,q)|$ as a function of doping density 
$\delta$ for $U=4$. The value of $t'$ is indicated by the shape of the symbol.
Here ${\vv p}=(-\pi,0)$, ${\vv q}=(\pi,\pi)$, and $\beta=4$.}
\label{Damp_Gam_chi}
\end{figure}

\begin{figure}
\centering
\epsfig{file=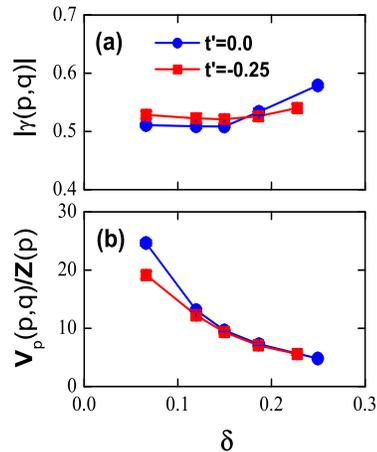,height=7cm,width=6cm,angle=0}
\caption{(color online)
(a) El-sp vertex $|\gamma(p,q)|$ and (b) $V_{p}(p,q)/Z(p)$
as a function of doping density $\delta$ for $U=4$.
${\vv p}$ and ${\vv q}$ are the same as in Fig.~\ref{Damp_Gam_chi}.}
\label{gamma_VZ_t1}
\end{figure}

As discussed in the Introduction, previous work on the spin-fermion
model~\cite{chub} suggests that the vertex correction
$|\Gamma-1|$ (not including $Z$)
gets considerably reduced, whenever spin fluctuations get damped.
To explore this fact,
we have compared numerical results for different values of the
next-nearest-neighbor hopping
($t'=0$ and $t'=-0.25$), since one expects the damping to increase for
larger negative $t'$. 
The spin susceptibility $\chi_{zz}$, a quantity proportional to 
the damping rate $\gamma_d$, 
and the el-sp vertex $\Gamma$ are plotted 
in Figs.~\ref{Damp_Gam_chi}(a)-\ref{Damp_Gam_chi}(c) as
a function of doping density. When $t'=0$, 
the Fermi surface is particlelike
and encloses the zone center $\Gamma=(0,0)$. On the other hand, 
when $t'=-0.25$, the Fermi 
surface is holelike and encloses the zone corner $Q=(\pi,\pi)$, and,
in particular, it crosses the magnetic Brillouin Zone boundary at
so-called hot spots.
In this second case, spin fluctuations get strongly damped due to the
decay into  particle and holes near the Fermi surface.
Due to the finite temperature and short correlation length, 
a damping of spin fluctuations is also
present for $t'=0$. The quantity $\gamma_d = b/c$
shown in Fig.~\ref{Damp_Gam_chi}, is obtained by fitting the spin 
susceptibility $\chi(q,i\omega_{m})^{-1}$ to the form
$a+b\omega_{m}+c\omega_{m}^{2}$, and is, thus, proportional to the 
spin damping rate.
In the underdoped region,
the spin susceptibility is strongly suppressed by a negative $t'$ 
due to its frustrating effect on the AF alignment, whereas 
the difference of $\gamma_d$ is small, as seen in 
Fig.~\ref{Damp_Gam_chi}(a) and Fig.~\ref{Damp_Gam_chi}(b).
On the other hand, in the overdoped region,
where the holelike Fermi surface passes
close to $(\pi,0)$ and symmetry points (whereas the particlelike
Fermi surface moves away from the magnetic Brillouin zone boundary),
$t'$ has little effect on the spin susceptibility, 
while the spin damping rate is dramatically increased by a
negative $t'$.
Fig.~\ref{Damp_Gam_chi}(c) clearly shows that the el-sp
vertex $|\Gamma|$ is larger for $t'=-0.25$ than for $t'=0$, 
demonstrating that the vertex correction is reduced
either by suppressing the spin susceptibility or by
increasing the spin damping rate. 
However, the largest difference in $|\Gamma|$ is not observed
at the smallest doping density $\delta=0.066$, where the suppression
of the spin susceptibility is strongest, or near the doping density 
$\delta=0.20$, where the difference of the damping rates is largest. 
This result suggests that, although both the spin susceptibility
and the damping of spin fluctuations are related to the 
magnitude of vertex corrections, they are probably not the only 
relevant factors.

In Figs.~\ref{gamma_VZ_t1}(a) and \ref{gamma_VZ_t1}(b) we show the effects
of the next-nearest-neighbor hopping $t'$ on the effective 
el-sp vertex $\gamma(p,q)$
and the effective pairing interaction
$V_{p}(p,q)/Z(p)$. In contrast to $\Gamma$, the vertex $\gamma \sim \Gamma/Z$
shows a weak dependence on  $t'$ (and also on $\delta$),
and in the overdoped regime is even smaller for 
$t'=-0.25$ than for $t'=0$. The different behavior of the vertices 
$\Gamma$ and $\gamma$ is obviously due to the $Z$ factor, which increases
with increasing $|t'|$ and/or decreasing doping.
The result fo $V_{p}(p,q)/Z(p)$ shows that the pairing strength 
has almost no dependence 
on $t'$ in the optimal and overdoped regimes, and is somewhat reduced 
at a negative $t'$ in the underdoped regime.

Finally, we  comment on the overall sign of the vertex correction.
In our cases studied, as shown in Fig.~\ref{Vp_Gamma_chi} and 
Fig.~\ref{Damp_Gam_chi}, $|\Gamma|$ is always less than $1$,
indicating that the el-sp interaction is suppressed by vertex corrections. 
This is in contrast to 
the finding in the spin-fermion model, which shows that the total
vertex correction at moderate doping has a positive sign, i.~e., 
vertex correction actually increases the el-sp interaction. 
This effect, however, is rather small and for moderate $\xi$
may be overshadowed by
the contributions from high-energies ($O(E_F)$), not included
in the low-energy spin-fermion models.  The comparison
between our calculations and the results in the spin-fermion model 
shows that, while the effect of damping of spin fluctuations on vertex
corrections is 
quite robust and agrees with low-energy considerations,
the sign and the magnitude 
of the vertex correction near optimal doping may be model dependent.

\section{\label{Conclusions} Conclusions}
In summary, based on Quantum-Monte-Carlo simulations, we have studied
the renormalization of the el-sp  vertex in the two-dimensional
$t-t'-U$ Hubbard model. We found that 
the fully renormalized el-sp vertex between bosons and free fermions 
decreases quite generally with decreasing temperature at all
spin-fluctuation momentum transfers.
We distinguished between vertex renormalization itself and 
the renormalization due to the 
dressing of the vertex by quasiparticle $Z$ factors.  
We analyzed the effect of adding a negative next-nearest-neighbor 
hopping term $t'$ to the dispersion. This term changes the topology of 
the Fermi surface and allows a  spin fluctuations to 
decay into an electron-hole pair.
We found that $t'$  reduces vertex corrections (but not $Z$),
in agreement with
previous results on the spin-fermion model. However, in contrast to
Ref.~\onlinecite{chub}, we did not observe 
a positive vertex correction,
i.~e. the renormalized vertex is always smaller than the bare one.

The suppression of the fully renormalized el-sp vertex, particularly 
at $t'=0$, gives rise to a
substantial reduction of the pairing mediated by antiferromagnetic spin
fluctuations in both the intermediate- and strong-correlation regimes. 
This result extends Schrieffer's argument (Eq.~\ref{gamma_sc})
about the suppression of the el-sp vertex to the case in which the AF
precursor is no longer present.
Notice, however, that  $\Gamma$ 
and $\chi$ behave in the opposite way (as implied by 
Eq.~\ref{gamma_sc}, only when one considers their 
{\em temperature} behavior.
In contrast, when {\em doping} decreases at a fixed temperature, 
$\Gamma$ increases with increasing $\chi$ (the results are not 
shown here). Thus, in the situation when the system is away 
from the AF precursor, there is no general proportionality relation 
 implied in Eq.~\ref{gamma_sc}. 

The authors thank D.J.~Scalapino for useful discussions.
The W\"urzburg group acknowledges support by the DFG under 
Grant No.~DFG-Forschergruppe 538, 
by the Bavaria California Technology Center (BaCaTeC), and the KONWHIR 
project CUHE. The work of Z.B.H was supported in part by the National 
Science Foundation Grant No. 10574040. 
EA was partly supported by the FWF project N. P18551-N16. 
AVC is supported by  NSF-DMR 0240238.
The calculations were carried out at the 
high-performance computing centers LRZ (M\"unchen) and
HLRS (Stuttgart).


\begin{thebibliography}{99}
\bibitem{note1}
Note that in Ref.~\onlinecite{schrieffer}, the renormalization of $Z$ was 
not included, i.e., the theoretical analysis assumed that $Z=1$,
and $\gamma =\Gamma$.
\bibitem{pq} 
In our convention, unbolded variables denote both
Matsubara frequency and momentum, i.~e., $p=(p_0,{\vv p})$ and
$q=(q_0,{\vv q})$. We have set the frequencies to their minimum 
values, i.~e., $p_0=\pi T$ for fermions and $q_0=0$ for bosons.
\end{thebibliography}
\end{document}